\documentclass[aps,twocolumn,superscriptaddress,showpacs,amsmath]{revtex4}

\usepackage{epsf,latexsym}
\usepackage{amssymb,amsmath}
\usepackage{times}

\newcommand{\bra}[1]{\langle #1 |}
\newcommand{\ket}[1]{| #1 \rangle}

\newcommand{\qed}{$\hfill \Box$}

\newcommand{\ignore}[1]{}

\newcommand{\be}{\begin{equation}}
\newcommand{\ee}{\end{equation}}

\def\CC{{\rm\kern.24em \vrule width.04em height1.46ex depth-.07ex
    \kern-.30em C}}
\def\P{{\rm I\kern-.25em P}}

\def\RR{{\rm
         \vrule width.04em height1.58ex depth-.0ex
         \kern-.04em R}}

\def\bbbone{{\mathchoice {\rm 1\mskip-4mu l} {\rm 1\mskip-4mu l}
{\rm 1\mskip-4.5mu l} {\rm 1\mskip-5mu l}}}

\def\bbbc{{\mathchoice {\setbox0=\hbox{$\displaystyle\rm C$}\hbox{\hbox
to0pt{\kern0.4\wd0\vrule height0.9\ht0\hss}\box0}}
{\setbox0=\hbox{$\textstyle\rm C$}\hbox{\hbox
to0pt{\kern0.4\wd0\vrule height0.9\ht0\hss}\box0}}
{\setbox0=\hbox{$\scriptstyle\rm C$}\hbox{\hbox
to0pt{\kern0.4\wd0\vrule height0.9\ht0\hss}\box0}}
{\setbox0=\hbox{$\scriptscriptstyle\rm C$}\hbox{\hbox
to0pt{\kern0.4\wd0\vrule height0.9\ht0\hss}\box0}}}}

\def\bbbz{{\mathchoice {\hbox{$\sf\textstyle Z\kern-0.4em Z$}}
{\hbox{$\sf\textstyle Z\kern-0.4em Z$}}
{\hbox{$\sf\scriptstyle Z\kern-0.3em Z$}}
{\hbox{$\sf\scriptscriptstyle Z\kern-0.2em Z$}}}}

\newcommand{\putfig}[2]{$$\leavevmode\hbox{\epsfxsize=#2 cm
   \epsffile{#1.eps}}$$}

\begin{document}
\title{Ground state entanglement and geometric entropy in the Kitaev model}
\author{Alioscia Hamma}
\affiliation{Institute for Scientific Interchange (ISI), Villa Gualino, Viale Settimio Severo 65, I-10133 Torino, Italy}
\affiliation{Dipartimento di Scienze Fisiche, Universit\`a Federico II, Via Cintia ed.~G, 80126 Napoli, Italy}

\author{Radu Ionicioiu}
\affiliation{Institute for Scientific Interchange (ISI), Villa Gualino, Viale Settimio Severo 65, I-10133 Torino, Italy}

\author{Paolo Zanardi}
\affiliation{Institute for Scientific Interchange (ISI), Villa Gualino, Viale Settimio Severo 65, I-10133 Torino, Italy}

\begin{abstract}
We study the entanglement properties of the ground state in Kitaev's model. This is a two-dimensional spin system with a torus topology and nontrivial four-body interactions between its spins. For a generic partition $(A,B)$ of the lattice we calculate analytically the von Neumann entropy of the reduced density matrix $\rho_A$ in the ground state. We prove that the geometric entropy associated with a region $A$ is linear in the length of its boundary. Moreover, we argue that entanglement can probe the topology of the system and reveal topological order. Finally, no partition has zero entanglement and we find the partition that maximizes the entanglement in the given ground state.
\end{abstract}
\maketitle

Entanglement is a topic of common interest in the two fields of quantum information and condensed matter theory. On one hand entanglement is the essential resource for quantum information processing and quantum communication. Protocols based on entangled states can offer an exponential speedup with respect to classical computation \cite{QC}. On the other hand, since entanglement measures quantum correlations for pure states, it is natural to think of a close connection between entanglement and the correlation functions of highly correlated states in condensed matter systems \cite{popp}. Entanglement seems to play an important role also in quantum phase transitions, where is believed to be responsible for the appearance of long-range correlations \cite{osborne,latorre}. Hence the quantification of entanglement in quantum states is a crucial problem for both fields.

In quantum physics, it is important to determine whether a system can be simulated classically or not. Indeed, a quantum system that can offer a true quantum computation cannot be efficiently simulated by a classical dynamics. Recent works have demonstrated that any slightly entangled quantum system, i.e., a system in which the entanglement scales less than logarithmically with the number of qubits, can be efficiently simulated by a classical computer \cite{vidal}.  It is henceforth important to determine how the entanglement scales with the size of the system. To this end, we can define the {\em geometric entropy} \cite{callan} associated with a pure state and a geometrical region $A$ as the von Neumann entropy of a reduced density matrix, which is the measure of entanglement between the degrees of freedom inside the region $A$ and outside it.  The scaling of entanglement with the size of the system is connected to the {\em holographic principle} \cite{holographic}, i.e., the hypothesis that the entropy of a region is proportional to the area of its boundary. In the case of a critical spin chain, instead, the entanglement between a spin block of size $L$ and the rest of the chain scales like $S\sim \log_2 L$ and thus this system can be simulated classically \cite{latorre,cardy}.

In this article we study the ground state entanglement in Kitaev's model \cite{kitaev}. This is a 2-dimensional exactly solvable spin system on a torus with four-body interactions in the Hamiltonian. Its relevance stems from the fact that it was the first example of the new subject of topological quantum computation \cite{kitaev,top} and because it features {\em topological order} \cite{wen}. This is a type of quantum order that can describe those states of matter that are not associated to symmetries, like the fractional quantum Hall liquids. Topological order is not related to the symmetries of the Hamiltonian and it is robust against arbitrary local perturbations, even those that destroy all the symmetries of the Hamiltonian. 

By considering a subsystem of spins in a region $A$, we shall prove that the geometric entropy of $A$ is linear in the length of its boundary. Moreover, we find that entanglement can probe the topology of the system and conjecture it can detect topological order.

{\em The model.} Consider a $k\times k$ square lattice on the torus with spins-$\frac{1}{2}$ attached to the links. There are $2k^2$ links and the total Hilbert space $\mathcal{H}$ has dimension $2^{2k^2}$. The geometrical objects of the model are stars and plaquettes. A {\em star} is the set of four links sharing a common vertex. A {\em plaquette} is an elementary square (face) on the lattice. For any star $s$ and plaquette $p$ we define the star operator $A_s$ and plaquette operator $B_p$ as: $A_s= \prod_{j\in s} \sigma^x_j  ,\  B_p= \prod_{j\in p}\sigma^z_j$. The Hamiltonian of the system is:
\begin{equation}
\label{hamiltonian}
H= -\sum_s A_s -\sum_p B_p
\end{equation}

These operators share either 0 or 2 links so they all commute: $[A_s,B_p]=0\ , \forall s,p$ and hence the model is exactly solvable. Its ground state is $\mathcal{L}=\{\ket{\xi}\in \mathcal{H}:A_s\ket{\xi}=B_p\ket{\xi}=\ket{\xi},\ \ \forall s,p\}$. We have the following two constraints on the stars and plaquettes: $\prod_{\forall s} A_s= \prod_{\forall p} B_p= \bbbone$, so not all the $A_s,B_p$ are independent (there are only $k^2-1$ independent operators of each type). Since they all commute, we can use them to label the states. By imposing the $2k^2-2$ independent constraints $A_s=B_p=1$ we can label $2^{2k^2-2}$ states in the ground state. Hence we find that the ground state is 4-degenerate, $\dim \mathcal{L}= 2^{2k^2-(2k^2-2)}=4$. More generally, the lattice does not need to be square. On a Riemann surface of genus $\mathfrak{g}$, the same model has a $4^\mathfrak{g}$-fold degenerate ground state, depending on the genus of the surface. This degeneracy is the sign of {\em topological order} \cite{wen}.

Given a curve $\gamma$ running on the links of the dual lattice, we define the associated string operator $W^x[\gamma]= \prod_{j\in\gamma} \sigma^x_j$; by $j\in\gamma$ we mean all the links crossed by the curve $\gamma$ connecting the centers of the plaquettes. A {\em string-net} is a product of string operators.

Let $\mathcal{A}$ be the group generated by the $k^2-1$ independent star operators $A_s$; $|{\mathcal A}|\equiv \mbox{card}\, {\mathcal A}= 2^{k^2-1}$. The elements $g\in {\mathcal A}$ are products of closed contractible strings \cite{kitaev,hiz}. Let $\gamma_1,\gamma_2$ be the two non-contractible loops generating the fundamental group of the torus $\pi_1(T^2)$ and define the associated string operators $w_{1,2}= W^x[\gamma_{1,2}]$. We refer to $w_i$ as ``ladder'' operators since they flip all the spins along a ladder going around a meridian, respectively a parallel, of the torus. Let $\overline{N}$ be the group generated by $\mathcal{A},w_1,w_2$. This group contains all the possible products, contractible or not, of closed strings. Closed strings and closed string-nets commute with $H$, while open strings do not commute \cite{kitaev}. The group $\overline{N}$ splits in four cosets with respect to $\mathcal{A}:\mathcal{A}e\equiv\overline{N}_{00},\mathcal{A}w_1\equiv\overline{N}_{01},\mathcal{A}w_2\equiv\overline{N}_{10},\mathcal{A}w_1w_2\equiv\overline{N}_{11}$. Since $\overline{N}$ is Abelian, right and left cosets coincide. The four cosets are the equivalence classes in $\overline{N}/\mathcal{A}$ and all have the same order $2^{k^2-1}$; hence $|\overline{N}|=2^{k^2+1}$.

Let $\mathcal{S}$ be an orthonormal basis of the Hilbert space $\mathcal{H}$, $\mathcal{S}=\{\ket{e}\equiv\ket{s_1,...,s_{2k^2}}:s_j=0,1\}$, where each $s_i$ labels the 2-dimensional Hilbert space of a single spin/link in the lattice.

Let $\ket{00}\equiv \ket{s_1=0,...,s_{2k^2}=0}$. Consider the orbit of $\ket{00}$ through $\overline{N}$,  $\mathcal{S}'\equiv \overline{N} \ket{00}=  \{ g\ket{00} |\ \  g\in\overline{N} \}$. Thus $\mathcal{S}'$ is the homogeneous space with respect to $\overline{N}$. Since the action of $\overline{N}$ on $\mathcal{S}$ is free, we have $|\mathcal{S}'|= 2^{k^2+1}$.

{\em The ground state.} We now give an explicit construction of the vectors $\ket{\xi}$ in the ground state $\mathcal{L}$. The constraint $B_p\ket{\xi}=\ket{\xi}, \forall p$ means that the ground state is a superposition of only basis vectors in $\mathcal{S}'$ \cite{kitaev}, $\ket{\xi}= \sum_{\ket{e_j}\in\mathcal{S}'} a_j \ket{e_j}= \sum_{g\in\overline{N}}a(g)g\ket{00}$.

The constraint $A_s\ket{\xi}=\ket{\xi}, \forall s$ implies that all the vectors from each sector $\overline{N}_{ij} \ket{00}$ are in the superposition with the same coefficient, that is $\ket{\xi}\in\mathcal{L}$ if and only if $\ket{\xi}= |\mathcal{A}|^{-1/2}\sum_{i,j=0}^1a_{ij}\sum_{g\in\overline{N}_{ij}}g 	\ket{00}$ with $\sum_{i,j=0}^1|a_{ij}|^2= 1$. This is because $A_s\sum_{g\in\overline{N}_{ij}} g\ket{00}= \sum_{g\in\overline{N}_{ij}} g\ket{00}$ since each sector is mapped into itself by star operators, since $A_s \in \overline{N}$, $A_s\overline{N}_{ij}= \overline{N}_{ij}$ and there are no invariant subspaces within $\overline{N}_{ij}$. We can thus write the ground state subspace as $\mathcal{L}= \mbox{span} \{\ket{\xi_{ij}},\  i, j=0,1\}$, where $\ket{\xi_{ij}}=|\mathcal{A}|^{-1/2}\sum_{g \in {\mathcal A}} g \ket{ij}$ and $\ket{ij}= w_1^j w_2^i \ket{00}$. By construction $\ket{\xi_{ij}}$ are orthonormal.

{\em Entanglement properties of the ground state.} Multipartite entanglement is notoriously difficult to quantify, since there is no known entanglement measure for the general state of a quantum system. However, for an arbitrary pure state $\rho_{AB}$ of a bipartite system $(A,B)$ there is an (essentially) unique entanglement measure, the von Neumann entropy $S$:
\begin{equation}
S\equiv -\mbox{Tr} (\rho_A \log_2 \rho_A)
\label{entropy}
\end{equation}
where $\rho_A= \mbox{Tr}_B (\rho_{AB})$ is the reduced density matrix of the sub-system $A$.

{\em Proposition 1.} For a given lattice partition $(A,B)$, the four ground states $\ket{\xi_{00}}, \ket{\xi_{01}}$, $\ket{\xi_{10}}, \ket{\xi_{11}}$ have the same entropy of entanglement $S$.

{\em Proof.} Let $w_i\equiv w_{i,A} \otimes w_{i,B}$, where $w_{i,A(B)}$ acts only on the $A$ ($B$) subsystem. From the circular property of the trace and $w^2_i= \bbbone$, $w_i^\dag= w_i$, $i= 1,2$, it follows immediately that all four ground states have isospectral reduced density matrices, e.g., $\rho_A (\xi_{01}) = \mbox{Tr}_B(\ket{\xi_{01}}\bra{\xi_{01}})= \mbox{Tr}_B(w_1\ket{\xi_{00}}\bra{\xi_{00}} w_1)= w_{1,A} \mbox{Tr}_B(\ket{\xi_{00}}\bra{\xi_{00}}) w_{1,A}$. Therefore $S$ is the same for all four ground states. \qed

Any element $g\in \overline{N}$ can be decomposed as $g= x_A \otimes x_B$, with $x_A (x_B)$ a product of spin flip operators $\prod_{j} \sigma^x_j$ acting only on the subsystem $A(B)$. In general $x_{A,B}$ are not necessarily closed string nets. If, however, $x_A$ is a closed string-net (from which follows immediately that also $x_B$ is), we will write $g= g_A \otimes g_B$, $g_{A,B}\in \overline{N}$.

We introduce now the following subgroups of $\mathcal{A}$ acting trivially on the subsystem $A$ and respectively $B$:
\begin{eqnarray}\label{aa}
\mathcal{A}_A &\equiv& \{g\in\mathcal{A}\ |\ \  g= g_A\otimes \bbbone_B\} \\
\mathcal{A}_B &\equiv& \{g\in\mathcal{A}\ |\ \  g= \bbbone_A\otimes g_B\}
\end{eqnarray}
The order of these groups is $d_A\equiv |\mathcal{A}_A|$ and $d_B\equiv |\mathcal{A}_B|$.  Define also the quotient $\mathcal{A}/\mathcal{A}_B$ and let $f=|\mathcal{A}/\mathcal{A}_B|= |\mathcal{A}|/d_B$ be its cardinality. Notice that $f$ is the number of elements in $\mathcal{A}$ that act freely on $A$. If there are $n$ independent stars $A_s$ acting on $A$, it turns out that $f=2^n$. In other words, $f$ is the number of different configurations of spins on $A$ compatible with the states in $\mathcal{S}'$. Define the group $\mathcal{A}_{AB}\equiv \mathcal{A}/(\mathcal{A}_{A}\cdot\mathcal{A}_{B})$. We have $|\mathcal{A}_{AB}|= |\mathcal{A}|/d_A d_B$. 

{\em Lemma.} The reduced density matrix corresponding to the basis vector $\ket{\xi_{00}}$  is $\rho_A = f^{-1} \sum_{\substack{g\in \mathcal{A}/\mathcal{A}_B\\ 
\tilde{g}\in\mathcal{A}_A}} x_A\ket{00_A}\bra{00_A}x_A\tilde{g}_A$.

{\em Proof.} First, recall that any two group elements $g,g'\in \mathcal{A}$ are related by $g'= g \tilde{g}$, with $\tilde{g}\in \mathcal{A}$. We can write the reduced density matrix for the state $\ket{\xi_{00}}$ as $\rho_A= |\mathcal{A}|^{-1}\, \mbox{Tr}_B\sum_{g,g'\in {\mathcal A}} g\ket{00}\bra{00}g'= |\mathcal{A}|^{-1}\sum_{g,g'\in {\mathcal A}}x_A\ket{00_A}\bra{00_A}x'_A\bra{00_B}x'_Bx_B\ket{00_B}$; the non-zero elements of this sum satisfy $x'_Bx_B=\bbbone$, which implies $x'_A= x_A\tilde{g}_A$, with $\tilde{g}=\tilde{g}_A\otimes \bbbone_B \in \mathcal{A}_A$. Therefore $\rho_A= |\mathcal{A}|^{-1}\sum_{g\in {\mathcal A}, \tilde{g}\in\mathcal{A}_A} x_A \ket{00_A}\bra{00_A} x_A \tilde{g}_A$. Now, $x'_A= x_A$ if and only if $g'= h g$, with $h\in\mathcal{A}_B$, and since $f^{-1}=d_B/|\mathcal{A}|$, we can write
\begin{eqnarray}
\rho_A = f^{-1} \sum_{\substack{g\in \mathcal{A}/\mathcal{A}_B\\ 
\tilde{g}\in\mathcal{A}_A}} x_A\ket{00_A}\bra{00_A}x_A\tilde{g}_A
\end{eqnarray}
Notice that if $d_A= 1$, the reduced density matrix is diagonal.\qed

We are now ready to prove the main result.

{\em Theorem.} Consider a partition $(A,B)$ of the lattice, and let the system be in the ground state $\ket{\xi_{00}}$. The entropy of entanglement is $S= \log_2 (f/d_A)= \log_2|\mathcal{A}|- \log_2(d_A d_B)= \log_2|\mathcal{A}_{AB}|$.

{\em Proof.} Let us compute the square of the reduced density matrix:
\begin{eqnarray}
\nonumber
\rho_A^2 &=& f^{-2}\sum_{ \substack{g,g'\in
\mathcal{A}/\mathcal{A}_B\\ \tilde{g},\tilde{g}'\in\mathcal{A}_A}}x_A\ket{00_A}\bra{00_A}x_A\tilde{g}_A x'_A\ket{00_A}\bra{00_A}x'_A\tilde{g}'_A\\
\nonumber
&=& f^{-2}\sum_{ \substack{g\in
\mathcal{A}/\mathcal{A}_B\\\tilde{g},\tilde{g}'\in\mathcal{A}_A}}x_A\ket{00_A}\bra{00_A}x_A\tilde{g}_A\tilde{g}'_A\\
&=& f^{-2}|\mathcal{A}_A|\sum_{ \substack{g\in
\mathcal{A}/\mathcal{A}_B\\\tilde{g}\in\mathcal{A}_A}}x_A\ket{00_A}\bra{00_A}x_A\tilde{g}_A = f^{-1}d_A\rho_A
\end{eqnarray}
From the Taylor expansion of the $\log$ we obtain $\log_2\rho_A= \rho_A f d_A^{-1} \log_2(d_A/f)$ and the entropy of entanglement is 
\begin{eqnarray}
S = \log_2(f d_A^{-1})= \log_2 \frac{|\mathcal{A}|}{d_A d_B} = \log_2|\mathcal{A}_{AB}|.
\end{eqnarray}

\begin{figure}
\putfig{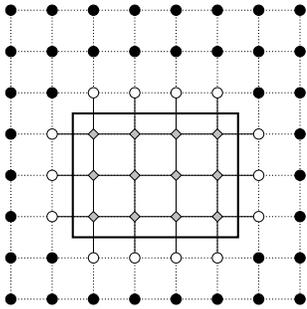}{4}
\caption{A region $A$ of the lattice obtained by taking all the spins (solid thin lines) inside or crossed by a convex loop (thick line). The length $L_{\partial A}$ of its boundary is equal to the number of spins intersected by the loop and this is equal to the number of (white) sites on which the stars act on both $A$ and $B$. The area $\Sigma_A$ of $A$ is equal to the number of (gray diamonds) sites inside it and $d_A= 2^{\Sigma_A}$. Then the area of $B$ (black sites) is $\Sigma_B= k^2-\Sigma_A- L_{\partial A}$. The geometric entropy is $S= L_{\partial A}- 1$.}
\label{bulk}
\end{figure}

The meaning of this formula is the following. Both $\mathcal{A}_A$ and $\mathcal{A}_B$ contain information related {\em exclusively} to the subsystems $A$ and $B$, respectively. Hence we say that it is the information contained in the {\em bulk}. Then the entropy is given by the difference between the total disorder and the disorder in the bulk. This implies that the entropy is due to the boundary between the two subsystems. If we choose the partitions in a convenient way, we can give a clear geometrical picture of this boundary. Consider now $A$ as the set of all spins inside or crossed by a loop in the dual lattice (see Fig.\ref{bulk}). The spins intersected by the loop are the {\em boundary} of $A$, while the ones inside are the {\em bulk}. The area $\Sigma_A$ of $A$ (in lattice units) is the number of sites inside the loop and $d_A= 2^{\Sigma_A}$, since this is the number of independent closed string-nets acting exclusively on $A$. Similarly $d_B= 2^{\Sigma_B}$, where $\Sigma_B$ is the number of star operators acting exclusively on $B$. Then the entropy is $S= \log_2(2^{k^2-1})- \log_2(d_A d_B)= k^2-1- \Sigma_A -\Sigma_B$. If the loop is {\em convex} (i.e., a rectangle), the perimeter $L_{\partial A}$ of $A$ is equal to the number of spins intersected by the loop and this is $L_{\partial A}= k^2- \Sigma_A- \Sigma_B$ (see Fig.\ref{bulk}). Therefore 
\begin{equation}
\label{loopentropy}
S= L_{\partial A}- 1
\end{equation}
Notice that in this case the entropy of a generic ground state $\ket{\xi}$ is also $S= L_{\partial A}- 1$. This follows from the fact that the ladder operators can always be moved outside the disk (see also \cite{hiz}).

If now we take as the boundary of $A$ an arbitrary (contractible) loop $\gamma$ on the lattice, the entropy will be $S= \alpha_1(\gamma)L_{\partial A}-1+\alpha_2(\gamma)$, where $\alpha_i(\gamma)$'s depend only on the geometry of the loop \cite{hiz}. Hence in general the entropy is linear in the boundary length of $A$. We see that this entropy is the {\em geometric entropy} \cite{callan} obtained by computing the pure state density matrix and tracing over the variables inside the geometric region $A$ to obtain a reduced density matrix to evaluate the entropy $S$. We want to explain the very important fact of the presence of the $-1$ in Eq.(\ref{loopentropy}). The $-1$ comes from the constraint that we have on the number of independent stars; this follows from the fact that each link is shared by two sites, hence $\prod_{\forall s} A_s= \bbbone$. This construction is not restricted to a square lattice. On a Riemann surface of genus $\mathfrak{g}$ the number of sites, links and plaquettes ($n_0,n_1$ and $n_2$, respectively) is constrained by the Euler's formula: $n_0-n_1+n_2=2 (1-\mathfrak{g})$. The number of independent stars and plaquettes is again $n_0-1$ and $n_2-1$ (the last constraint follows from the boundary conditions); hence we can label $n_0+n_2-2=n_1-2\mathfrak{g}$ states and thus the ground state is $2^{2\mathfrak{g}}$-fold degenerate. The system exhibits topological order \cite{wen}. Now we can sketch an argument to compute the entropy for the surface of genus $\mathfrak{g}$. If the number of stars and plaquette is the same ($n_0=n_2$), then the entropy is $S= n_1/2-\mathfrak{g}-\Sigma_A-\Sigma_B$. The {\em topological order} in this model manifests itself in both the degeneracy and the entanglement in the ground state. This suggests the very appealing possibility that entanglement could detect topological order.

In the following we calculate the entropy $S$ for several subsystems $A$ of the lattice. From Proposition 1, $S$ is the same for all the states $\ket{\xi_{ij}}$, hence we will consider the system to be in the $\ket{\xi_{00}}$ ground state. For a generic ground state $\ket{\xi}$ the results will be presented elsewhere \cite{hiz}.

{\em Example 0: one spin}. If the subsystem $A$ is a single spin, no closed string net can act just on it; hence $d_A=1$, $f=2$ and the entropy is $S= 1$. This shows that any spin is maximally entangled with the rest of the system.

{\em Example 1: two spins}. We now compute the entanglement between two arbitrary spins on the lattice. In this case we use as an entanglement measure the concurrence $C$. For a general mixed state $\rho_{ij}$ of two qubits, the concurrence is $C= \max \{0, \sqrt{\lambda_1}-\sqrt{\lambda_2}-\sqrt{\lambda_3}-\sqrt{\lambda_4} \}$, where $\lambda_1,\lambda_2,\lambda_3,\lambda_4$ are the eigenvalues (in decreasing order) of the matrix $\rho_{ij} (\sigma^y \otimes \sigma^y) \rho^*_{ij} (\sigma^y \otimes \sigma^y)$ \cite{wooters}. It is easy to see that no $g\in\mathcal{A}$ acts exclusively on the two spins, so $d_A=1$ and the reduced density matrix $\rho_{ij}$ is diagonal (we considered the two spins as the $A$ subsystem). Let $\rho_{ij}= \mbox{diag}(a,b,c,d)$; in follows immediately that $C= 0$ always, hence there is no two-qubit entanglement between any two spins.

{\em Example 2: the spin chain}. We take the partition $A$ consisting of all the spins belonging to a meridian (or parallel) $\gamma_1$ of the torus; this is a system of $k$ spins-$\frac{1}{2}$. It can be easily seen that no closed string-net can act exclusively on $A$ and thus $d_A= 1$. The number of possible configurations of spins on the chain $\gamma_1$ is $2^k$, but there are only $f=2^{k-1}$ configurations of spins in $A$ that enter the ground state, namely the ones with an even number of spin flips. We see indeed that we have $k-1$ stars acting independently on the chain and we can obtain all the allowed configurations applying products of these stars (i.e., elements in $\mathcal{A}$ acting freely on the chain), which gives $f=2^{k-1}$. Then the entropy is $S= \log_2 f= k-1$.

{\em Example 3:  the ``vertical spins''}. Let $A$ be the set of all vertical spins on the lattice (and hence $B$ is the set of all horizontal spins). This partition is such that no closed string operator can act trivially on either subsystem, hence $d_A= d_B=1$ and we obtain the maximum possible entanglement $S=k^2-1$. 

It is interesting that no partition has zero entanglement, since $d_A d_B<|\mathcal{A}|$ for any (nontrivial) partition of the lattice. 

{\em Conclusions.} Entanglement is one of the most striking features of quantum mechanics and the most important resource for quantum information processing. Entanglement between two parts of a quantum system is measured by the von Neumann entropy. The entropy of entanglement can be connected, with some assumptions \cite{landau}, to the notion of entropy in thermodynamics. A whole number of studies is devoted to investigate how the entropy associated with a region is related to the geometric properties of that region. A simple argument by Srednicki \cite{srednicki} for a free massless scalar field says that entropy should scale like the area of a spherical surface enclosing the field. This is because the entropy associated with the volume inside the sphere must be equal to the entropy associated with the volume {\em outside the sphere}. Then the entropy should depend only on properties that are shared by the two regions (inside and outside the sphere), that is, the area of the shared boundary.  Srednicki \cite{srednicki} has calculated (numerically) the entropy for $2+1$ and $3+1$ conformal field theories. It turns out that the entropy associated with a region $\mathcal{R}$ is proportional to the size $\sigma(\mathcal{R})$ of its boundary. This result is in agreement with black-hole thermodynamics, whose entropy is $S_{BH}= M^2_{Pl}A/4$, where $M_{Pl}$ is the Planck mass and $A$ is the surface area of the horizon of the black hole \cite{beckenstein}. A recent work \cite{plenio} proves analytically that the entropy associated with a region of a discretized free quantum Klein-Gordon field can be bounded from above and below by quantities proportional to the surface area. Several authors have calculated the entanglement in 1D spin chains. In the case of $XY$ and Heisenberg models, the authors in Refs.~\cite{latorre} have calculated the entanglement between a spin block of size $L$ and the rest of the chain. They found two characteristic behaviors. For critical spin chains, the entanglement scales like $S\sim \log_2 L$, whereas for the noncritical case $S$ saturates with the size $L$ of the block. This result is in agreement with the result for black-hole thermodynamics in 1+1 dimensions \cite{larsen,preskill}, which suggested a connection between the entanglement measured in quantum information and the entropy of the vacuum in quantum field theories.

In this article we have investigated the ground state entanglement in the Kitaev's model. This is a two-dimensional spin system with four-body spin-spin interaction which presents topological order. On a Riemann surface of genus $\mathfrak{g}$ the degeneracy of the ground state is $4^{\mathfrak{g}}$ and such degeneracy is stable against local perturbations. This is at the root of the topological quantum computation. We found that although no two spins of the lattice are entangled (the concurrence is zero for any two spins), the ground state has multi-body entanglement. For a generic partition $(A,B)$ of the lattice we calculated analytically the von Neumann entropy of the reduced density matrix $\rho_A$ in the ground state $\ket{\xi_{ij}}$. We found that the geometric entropy associated with a region $A$ is linear in the length of its boundary. Moreover, no partition gives zero entanglement, so the system has an {\em absolute entanglement}. Finally, we argued that entanglement can probe the topology of the system and raised the very interesting question of whether it could detect (or measure) topological order. 

P.Z.~gratefully acknowledges funding by European Union project TOPQIP (contract IST-2001-39215).

\noindent {\em Note.} After the completion of this work, we came across Ref.~\cite{fattal} which uses a similar formalism in the context of stabilizer states.

%%%%%%%%%%%%%%%%%%%%%%%%%%%%%%%%%%%%%%%%%%%%%%%%%%%%%%%%%%%%%%%%%%%%%%%%%%%%%%%%%%%%%%%%%%%%%%%%%%%%

\end{document}